\documentclass[letterpaper, 10 pt, conference]{ieeeconf}  

\IEEEoverridecommandlockouts                              

\usepackage[english]{babel}
\usepackage{amsmath, amssymb, amsfonts, bm}
\usepackage{mathrsfs}
\usepackage{textcomp}
\usepackage[mathcal]{euscript}
\usepackage{enumerate}
\usepackage{mathbbol}
\usepackage{float}
\usepackage{color,graphicx,epstopdf}
\usepackage{dsfont}
\usepackage{hyperref}
\usepackage[noadjust]{cite}
\usepackage{caption}
\usepackage{subcaption}
\usepackage{mathtools}
\usepackage{tikz-cd}
\usepackage{mathtools}
\usepackage{algorithm}
\usepackage{algpseudocode}
\usepackage{url}
\usepackage{authblk}
\usepackage{stackengine}

\DeclareMathOperator*{\argmin}{arg\,min}

\newtheorem{theorem}{Theorem}
\newtheorem{definition}{Definition}

\newtheorem{lemma}{Lemma}
\newtheorem{remark}{Remark}
\newtheorem{proposition}{Proposition}
\newtheorem{assumption}{Assumption}

\newtheorem{experiment}{Experiment}
\usepackage[left=1.5cm,right=1.5cm,top=2cm,bottom=2cm]{geometry}
\usepackage{float}

\title{\LARGE \bf
	Transactive Multi-Agent Systems over Flow Networks
}

\author{Yijun Chen$^{1}$, Zeinab Salehi$^{2}$, Elizabeth L. Ratnam$^{2}$, Ian R. Petersen$^{2}$, Guodong Shi$^{1}$
	\thanks{$^{1}$Australian Center for Field Robotics, School of Aerospace, Mechanical and Mechatronic Engineering, The University of Sydney, Sydney, Australia, email: \{yijun.chen, guodong.shi\}@sydney.edu.au}%
	\thanks{$^{2}$The Research School of Engineering, The Australian National University, Canberra, Australia, email: \{zeinab.salehi, elizabeth.ratnam, ian.petersen\}@anu.edu.au}%
}

\begin{document}

	\maketitle
	\thispagestyle{empty}
	\pagestyle{empty}
	
	\begin{abstract}
		This paper presented insights into the implementation of transactive multi-agent systems over flow networks where local resources are decentralized. Agents have local resource demand and supply, and are interconnected through a flow network to support the sharing of local resources while respecting restricted sharing/flow capacity. We first establish a competitive market with a pricing mechanism that internalizes flow capacity constraints into agents' private decisions. We then demonstrate through duality theory that competitive equilibrium and social welfare equilibrium exist and agree under convexity assumptions, indicating the efficiency of the pricing mechanism. Additionally, a new social acceptance sharing problem is defined to investigate homogeneous pricing when the optimal sharing prices at all agents under competitive equilibrium are always equal for social acceptance. A conceptual computation method is proposed, prescribing a class of socially admissible utility functions to solve the social acceptance problem. A special case of linear-quadratic multi-agent systems over undirected star graphs is provided as a pedagogical example of how to explicitly prescribe socially admissible utility functions. Finally, extensive experiments are provided to validate the results.
	\end{abstract}
	
	\section{Introduction}
	
	Future technologies are being structured as networked multi-agent systems that take advantage of the Internet of Things to support critical infrastructure systems such as energy and power grids, communication networks, and automotive transportation \cite{mesbahi2010graph}. Multi-agent systems refer to systems with a group of agents that hold their own decisions and preferences and interact with each other to achieve a common goal, such as resource allocation \cite{nedic2010constrained} and control coordination \cite{meng2013event}. Recently, transactive control has emerged as a new technology and been applied in various applications such as transactive energy \cite{nunna2017multiagent} and smart city transportation \cite{annaswamy2016emerging} as a market-based coordination to achieve certain system-level objectives \cite{li2020transactive}. 
	
	
	Effective resource allocation is a major focus for transactive multi-agent systems \cite{nair2018multi}. The goal of transactive multi-agent systems is to translate market coordination into local decisions that lead to optimal individual payoffs while maintaining system-level optimality \cite{salehi2021social}. In light of classical welfare economics theory, the careful pricing of the resource unit can possibly contribute to achieving this goal \cite{pindyck2018microeconomics}. The concept of social welfare equilibrium describes the optimality at the system level, where the overall agent utilities are maximized while taking network-wide demand/supply balance into account. The idea of competitive equilibrium describes individual optimality, where each agent maximizes its own payoffs, as the sum of utility from resource consumption and income/cost from resource trading, while overall supply and demand are balanced.

	There have been many research efforts in developing pricing mechanisms \cite{chen2017demand,chen2010two,conejo2010real}. To improve reliability and sustainability, a critical factor in the practical implementation of resource allocation in distribution networks and energy markets is voltage operation constraints \cite{kang2022event}.  Some work presented market mechanisms focusing on voltage regulation \cite{li2015market,sampath2022voltage,umer2023novel,liu2022fully}. For example, the work of \cite{li2015market} presented a market mechanism for a radial distribution network that internalized voltage constraints into the trading transaction and introduced effective trading rules.

	Another key factor to consider in the practical implementation of resource allocation is line capacity constraints that may limit the amount of resources that can be shared between two agents. 
	For example, in inventory sharing, multiple wholesalers hold inventory of a particular product. Resource allocation can be used to optimize the distribution of inventory among wholesalers. But there may be capacity constraints on how much inventory can be shared between two wholesalers. A wholesaler may only be able to share a limited amount of inventory with his partners due to logistics limitations and contract stipulations from manufacturers. This example highlights the importance of considering capacity limitations in the design and implementation of resource allocation for transactive multi-agent systems. Most existing market designs in the presence of line congestion were under the setting of transmission networks that are coupled with other transmission laws such as Kirchoff's law \cite{chao1996market,doukas2011electric,gonen2011electrical}. However, transmission laws do not exist in multi-agent systems like inventory-sharing systems.
	
	In this paper,  we are motivated to investigate transactive multi-agent systems operating over flow networks where resources are decentralized. Agents have local resource demand and supply, and are interconnected through a flow network to facilitate the sharing of local resources. There are only two constraints arising from flow networks. For each agent, the amount of trading must be equal to the net incoming and outgoing flow; the flow between any two agents is restricted.  Then, with the presence of flow constraints, the optimal sharing prices may be different at different agents due to punishment of flow reaching flow capacity boundary, which raises fairness problem and disadvantage certain agents. To improve the fairness, it inspires us to define a new social acceptance sharing problem. The main contributions of the paper are as follows:
	\begin{itemize}
		\item Inspired by \cite{li2015market},  we propose a competitive market with a pricing mechanism that flow capacity constraints are internalized into agents' private decisions, where optimal trading prices may be different at different agents.
		\item Using duality theory, we show that under convexity assumptions, competitive equilibrium and social welfare equilibrium exist and agree, revealing the efficiency of the pricing mechanism.
		\item We define a new social acceptance sharing problem to  investigate homogeneous pricing when the optimal sharing prices at all agents under the competitive equilibrium are always equal to each other for social acceptance. We propose a conceptual computation method, which focuses on prescribing a class of socially admissible utility functions. The social acceptance problem can be solved as long as agents select their utility functions from the prescribed class.
		\item A special case linear-quadratic multi-agent systems over undirected star graphs is provided, which serves as an pedagogical example regarding how to explicitly prescribe a class of socially admissible utility functions.
	\end{itemize}
	
	The rest of the paper is organized as follows. Section~\ref{sec:TMAS-FN} presents multi-agent systems over flow networks and proposes a competitive market with pricing mechanisms. Section~\ref{sec:equivalence} presents the efficiency of the competitive market. Section~\ref{sec:homogeneous_pricing} investigates homogeneous pricing, defines a new social acceptance sharing problem for multi-agent systems over flow networks and proposes an algorithm for conceptually prescribing a family of socially admissible utility functions under which the optimal sharing prices at all agents are always equal to each other. Section~\ref{sec:examples} conducts comprehensive numerical experiments. Section~\ref{sec:conclusion} concludes the paper.
	
	\section{Transactive Multi-agent Systems over Flow Networks (TMAS-FN)} \label{sec:TMAS-FN}
	In this section, we present multi-agent systems over flow networks where flow constraints are taken into account.
	\subsection{Resource Allocation in Multi-agent Systems (MAS)} 
	Consider  a multi-agent system consisting of $n$ agents indexed in the set $\mathcal N=\{1, 2, ..., n \}$. Each agent $i$ generates/holds $a_i \in \mathbb{R}^{\geq 0}$ units of local resources. The system resource capacity is defined as the total amount of resources available in the system, i.e., $C= \sum_{i=1}^n a_i$. Each agent $i$ makes a consumption decision to consume $x_i\in \mathbb{R}^{\geq 0}$ units of resources. The utility function related to agent $i$ consuming $x_i$ amount of resource is $f_i(x_i):\mathbb{R}^{\geq 0} \to \mathbb{R}$. In the meanwhile, agents are connected within a network to share their local resources through pricing mechanisms. Each agent $i$ further makes a trading decision to trade  $e_i \in \mathbb{R}$ units of resources with other agents. A physical constraint for each agent's trading decision is $e_{i} \leq a_{i} - x_{i}$. The price for unit resource exchange at each agent $i$ is denoted by $\lambda_{i} \in \mathbb{R}$. As a result, each agent $i$'s payoff is the summation of utility from consumption and income/cost from trading.
	
	Denote $\mathbf{a}=(a_1,\dots, a_n)^\top$ as the local resource profile,  $\mathbf{x}=(x_1,\dots, x_n)^\top$ as the resource consumption profile, $\mathbf{e}=(e_1,\dots,e_n)^\top$ as the traded resource profile and $\bm{\lambda} =  (\lambda_1,\dots,\lambda_n)^\top$ as the unit resource price profile.

	\subsection{Trading over Flow Networks (FN)}
	
	Consider a directed network $\mathcal{G} = (\mathcal{N},\mathcal{A})$ with a set $\mathcal{N}$ of $n$ nodes and a set  $\mathcal{A}$ of $m$ directed arcs. The directed flow network considered in this paper is a connected graph. Each arc $(i,j)$ starts from node $i$ and points to node $j$, where node $i$ is called tail and node $j$ is called head. The node-node adjacent matrix is denoted by $\mathbf{G} \in \mathbb{R}^{n\times n}$, where the $ij$-entry is $1$ if $(i,j) \in \mathcal{A}$ and is $0$ otherwise.  The node-arc incidence matrix is denoted by $\mathbf{A} \in \mathbb{R}^{n\times m}$. Each row of $\mathbf{A}$ corresponds to a node; and each column of $\mathbf{A}$ corresponds to an arc. The column corresponding to arc $(i,j)$ has two non-zero entries with $1$ at row $i \in \mathcal{N}$ and $-1$ at row $j \in \mathcal{N}$. 
	Denote by $\mathbf{A}_{st}$ the $st$-entry of matrix $\mathbf A$. We define $\mathbf{A}^{+} \in \mathbb{R}^{n\times m}$ such that for $s = 1, \dots,n$ and $t = 1, \dots, m$, we have $\mathbf{A}_{st}^{+} = 1$ if $\mathbf{A}_{st} = 1$; otherwise, $\mathbf{A}_{st}^{+} = 0$. Similarly, we define $\mathbf{A}^{-} \in \mathbb{R}^{n\times m}$ such that  for $s = 1, \dots,n$, and $t = 1, \dots, m$, we have $\mathbf{A}_{st}^{-} = -1$ if $\mathbf{A}_{st} = -1$; otherwise, $\mathbf{A}_{st}^{-} = 0$.
	
	Each arc $(i,j) \in \mathcal{A}$ has an associated positive capacity $u_{ij} > 0$ representing the maximum amount that can flow on arc $(i,j)$. The actual amount of flow that passes through arc $(i,j)$ is denoted by $y_{ij} \geq 0$, which cannot exceed the capacity of its arc, i.e., $y_{ij} \leq u_{ij}$.  For each node $i \in \mathcal{N}$, we define the supply/demand function $b(i):\mathcal{N} \to \mathbb{R}$ as the net actual flow of node $i$, which is described by 
	$	b(i) = \sum_{j:(i,j) \in \mathcal{A}}y_{ij} - \sum_{j:(j,i) \in \mathcal{A}}y_{ji}.$
	If $b(i) >0$, node $i$ is a supply node; if $b(i) <0$, node $i$ is a demand node; and if $b(i) = 0$, node $i$ is a transshipment node.  For a multi-agent system with trading decisions whose realization is over a flow network, the trading decision  $e_{i}$ at each agent $ i \in \mathcal{N}$ is a local supply/demand variable, which is equivalent to $b(i), i \in \mathcal{N}$ in flow networks. 
	
	Denote $\mathbf{y} = (y_{ij})_{(i,j) \in \mathcal{A}} \in \mathbb{R}^{m}$ and $\mathbf{u} = (u_{ij})_{(i,j) \in \mathcal{A}} \in \mathbb{R}^{m}$. For $k = 1,\dots,m$, denote the $k$th entry of $\mathbf{y}$ and $\mathbf{u}$ by $\mathbf{y}_{k} \in \mathbb{R}$ and $\mathbf{u}_{k} \in \mathbb{R}$, respectively. The supply/demand of each agent $i \in \mathcal{N}$ can be rewritten as $e_{i} = \sum_{k = 1}^{m} \mathbf{A}_{ik}\mathbf{y}_{k}.$ 
	\subsection{Trading and Pricing Mechanism}
	In this subsection, we propose a decentralized resource market mechanism for transactive multi-agent systems that respects the flow network constraints. For a transactive multi-agent system over a flow network, the proposed mechanism can internalize the flow limitations  into the local decisions of agents. 
	
	Denote $\beta^{\ast} \in \mathbb{R}$, $\mathbf{q}^{\ast} = [q_{1}^{\ast},\dots,q_{n}^{\ast}]^{\top} \in \mathbb{R}^{n}$, $\bm{\lambda} = [\lambda_{1}^{\ast},\dots,\lambda_{n}^{\ast}]^{\top} \in \mathbb{R}^{n}$ and $\bm{\xi} = [\xi_{1}^{\ast},\dots,\xi_{m}^{\ast}]^{\top} \in \mathbb{R}^{m}$.
	We now definite the notion of competitive equilibrium for TMAS-FN in Definition~\ref{def:ce}.
	\begin{definition}[Competitive Equilibrium]\label{def:ce}
		A competitive equilibrium $(\mathbf{x}^{\ast},\mathbf{e}^{\ast},\mathbf{y}^{\ast},\beta^{\ast}, \mathbf{q}^{\ast}, \bm{\xi}^{\ast},\bm{\lambda}^{\ast})$ for a TMAS-FN is achieved if the following conditions hold:
		\begin{itemize}
			\item [(i)] For each agent $i \in \mathcal{N}$, the pair $(x_{i}^{\ast},e_{i}^{\ast})$ solves the following maximization problem
			\begin{subequations}\label{eq:ce_max}
				\begin{align}
					\max_{x_{i},e_{i}} \quad  & f_{i}(x_{i}) + \lambda^{\ast}_{i}e_{i} \label{eq:ce_obj}\\
					\text{\rm s.t.} \quad & e_{i} \leq a_{i} - x_{i}, \label{eq:ce_e_x_a}\\
					\quad &  x_{i} \geq 0.
				\end{align}
			\end{subequations}
			\item [(ii)] The price for each agent $i$ satisfies 
			\begin{equation}\label{eq:ce_price}
				\lambda_{i}^{\ast} = - ( \beta^{\ast} + q_{i}^{\ast}), \ i=1,\dots,n.
			\end{equation}
			
			\item [(iii)] The trading decisions balance the total traded resource across the network, that is 
			\begin{equation}\label{eq:supply/demand_balance}
				\sum_{i=1}^{n} e_{i}^{\ast} = 0.
			\end{equation}
			
			\item [(iv)] The sum of incoming and outgoing flows of agent $i$ should be equal to the amount of trading at each agent $i \in \mathcal{N}$; that is,
			\begin{subequations}
				\begin{align}\label{eq:net_flow}
					e_{i}^{\ast} &= \sum_{k = 1}^{m} \mathbf{A}_{ik}\mathbf{y}_{k}^{\ast},  \  i = 1,\dots,n,\\
					\mathbf{y}_{k}^{\ast} &\geq 0, \  k = 1, \dots, m.
				\end{align}
			\end{subequations}
			
			\item [(v)] If the flow capacity constraints are not binding, the price for flow is zero:
			\begin{align}\label{eq:ce_capacity_dual}
				\xi_{k}^{\ast}(\mathbf{y}_{k}^{\ast} - \mathbf{u}_{k})& = 0, k =1,\dots,m.
			\end{align}
			
			\item [(vi)] There holds 
			\begin{equation}\label{eq:ce_y_stationary}
				\xi_{k}^{\ast} - \sum_{i=1}^{n}q_{i}^{\ast}\mathbf{A}_{ik}=0, k =1,\dots,m. 
			\end{equation}
		\end{itemize}
	\end{definition}
	
	We next assume that there is a social planner who is responsible for making decisions regarding the consumption decisions $x_i, i \in \mathcal{N}$ and trading decisions $e_i, i \in \mathcal{N}$ of all agents in the TMAS-FN, as well as determining resource flows $y_{ij}, (i,j) \in \mathcal{A}$ over the flow network. The social planner considers the social welfare maximization problem over the flow network. We present the notion of social welfare equilibrium in Definition~\ref{def:swe}.
	\begin{definition}[Social Welfare Equilibrium]\label{def:swe}
		The social welfare equilibrium $(\mathbf{x}^{\star},\mathbf{e}^{\star},\mathbf{y}^{\star})$ is achieved for a TMAS-FN when $(\mathbf{x}^{\star},\mathbf{e}^{\star},\mathbf{y}^{\star})$ maximizes the following optimization problem
		\begin{subequations}\label{eq:swe}
			\begin{align}
				\max_{\mathbf{x},\mathbf{e},\mathbf{y}} \quad &
				\sum_{i = 1}^{n} f_{i}(x_{i})\\
				\text{\rm s.t.} \quad & \sum_{i=1}^{n} e_{i} =0, \label{eq:swe_balance}\\
				\quad & e_{i} \leq a_{i} - x_{i},  \  i = 1,\dots,n, \label{eq:swe_e_x_a}\\
				\quad &e_{i} = \sum_{k = 1}^{m} \mathbf{A}_{ik}\mathbf{y}_{k},  \  i = 1,\dots,n, \label{eq:swe_flow}\\
				\quad & \mathbf{y}_{k} \leq \mathbf{u}_{k}, \ k = 1,\dots,m, \label{eq:swe_capacity_constraints}\\
				\quad & x_{i} \in \mathbb{R}^{\geq 0}, \  i = 1,\dots,n,\\
				\quad & \mathbf{y}_{i} \in \mathbb{R}^{\geq0},\  i = 1,\dots,m. \label{eq:swe_positive_y}
			\end{align}
		\end{subequations}
	\end{definition}
	The social welfare equilibrium describes the optimality from the system-level perspective.
	
	
	\subsection{Related Work}
	
	This work of transactive multi-agent systems over flow networks builds upon our previous work \cite{chen2022competitive,salehi2021social,salehi2022social}. The idea of imposing flow/line capacity constraints is orginated from the work of \cite{chao1996market,li2015market}. Different from line capacity constraints imposed in our work,  the work of \cite{li2015market} presented a market framework in the presence of voltage operation constraints. In \cite{chao1996market}, their framework considered line constraints for electricity transmission networks where other transmission laws such as Kirchoff's law are also present. However,  our work only considers two constraints arising from flow networks. For each agent, the amount of trading must be equal to the net flow amount going in and out; the flow between ant two agents is restricted.

	\section{Efficiency of Pricing Mechanisms}\label{sec:equivalence}
	In this section, we show that the competitive equilibrium(s)  and social welfare equilibrium(s) coincide with each other for TMAS-FN with concave utility functions.
	\medskip
	
	\begin{theorem}\label{thm:equivalence}
		Consider a TMAS-FN. Suppose each $f_{i}(\cdot)$ is concave over the domain $\mathbb{R}^{\geq 0}$. Then the social welfare equilibrium(s) and the competitive equilibrium(s) coincide. To be precise, the following statements hold:
		\begin{itemize}
			\item [(i)] if $(\mathbf{x}^{\star},\mathbf{e}^{\star},\mathbf{y}^{\star})$ is a social welfare equilibrium, then there exists $\beta^{\ast} \in \mathbb{R}, \mathbf{q}^{\ast} \in \mathbb{R}^{n}, \bm{\xi}^{\ast} \in (\mathbb{R}^{\geq 0})^{m},\bm{\lambda}^{\ast} \in \mathbb{R}^{n}$ such that $(\mathbf{x}^{\star},\mathbf{e}^{\star},\mathbf{y}^{\star},\beta^{\ast}, \mathbf{q}^{\ast}, \bm{\xi}^{\ast},\bm{\lambda}^{\ast})$ is a competitive equilibrium.
			
			\item [(ii)] if $(\mathbf{x}^{\ast},\mathbf{e}^{\ast},\mathbf{y}^{\ast},\beta^{\ast}, \mathbf{q}^{\ast}, \bm{\xi}^{\ast},\bm{\lambda}^{\ast})$ is a competitive equilibrium, then $(\mathbf{x}^{\ast},\mathbf{e}^{\ast},\mathbf{y}^{\ast})$ is a social welfare equilibrium. 
		\end{itemize}
	\end{theorem}
	
	{\it \noindent Proof:} (i) For each $i \in \mathcal{N}$, we define $\mathbb{X}_{i} = \{(x_{i},e_{i})|e_{i} \leq a_{i} - x_{i};x_{i} \geq 0\}$. Define $\mathbb{X} = \{(\mathbf{x},\mathbf{e})|(x_{i},e_{i}) \in \mathbb{X}_{i}, i \in \mathcal{N}\}$. Clearly, $\mathbb{X}$ is a polyhedral set. For any $(\mathbf{x},\mathbf{e},\mathbf{y})$ such that $(x_{i},e_{i}) \in \mathbb{X}_{i}$, the Lagrangian associated with~\eqref{eq:swe} is
	\begin{subequations}\label{eq:lagrangian}
		\footnotesize
		\begin{align}
			&L(\mathbf{x},\mathbf{e},\mathbf{y}, \beta, \mathbf{q}, \bm{\xi}) \notag\\ 
			=& -\sum_{i=1}^{n} f_{i}(x_{i}) + \beta \sum_{i=1}^{n}e_{i} + \sum_{i=1}^{n}q_{i}(e_{i} - \sum_{k = 1}^{m} \mathbf{A}_{ik}\mathbf{y}_{k}) + \sum_{k=1}^{m} \xi_{k}(\mathbf{y}_{k} - \mathbf{u}_{k})\\
			=& -\sum_{i=1}^{n} f_{i}(x_{i})  +\sum_{i=1}^{n} (\beta+q_{i})e_{i} + \sum_{k=1}^{m}(\xi_{k} - \sum_{i=1}^{n}q_{i}\mathbf{A}_{ik})\mathbf{y}_{k} - \sum_{k=1}^{m}\xi_{k}\mathbf{u}_{k}. \label{eq:xe_y_sperate}
		\end{align}
	\end{subequations}
	\noindent We define $$L^{\ast}( \beta, \mathbf{q}, \bm{\xi}) = \min_{(\mathbf{x},\mathbf{e}) \in \mathbb{X}, \ \mathbf{y} \in (\mathbb{R}^{\geq 0})^{m}}L(\mathbf{x},\mathbf{e},\mathbf{y}, \beta, \mathbf{q}, \bm{\xi}).$$ 
	If $(\beta^{\ast}, \mathbf{q}^{\ast}, \bm{\xi}^{\ast})$ are dual optimal (i.e., $(\beta^{\ast}, \mathbf{q}^{\ast}, \bm{\xi}^{\ast})$ $\in \arg \max L^{\ast}( \beta, \mathbf{q}, \bm{\xi})$), there holds from strong duality that
	\begin{equation}
		(\mathbf{x}^{\star},\mathbf{e}^{\star},\mathbf{y}^{\star}) \in \argmin_{(\mathbf{x},\mathbf{e}) \in \mathbb{X}, \  \mathbf{y} \in  (\mathbb{R}^{\geq 0})^{m}} L(\mathbf{x},\mathbf{e},\mathbf{y}, \beta^{\ast}, \mathbf{q}^{\ast}, \bm{\xi}^{\ast}).
	\end{equation}
	We know $(\mathbf{x}^{\star},\mathbf{e}^{\star})$ and $\mathbf{y}^{\star}$ are independant in~\eqref{eq:xe_y_sperate}, which leads to 
	\begin{subequations}
		\begin{align}\label{eq:strong_duality_xe}
			(\mathbf{x}^{\star},\mathbf{e}^{\star}) & \in \argmin_{(\mathbf{x},\mathbf{e}) \in  \mathbb{X}} \Big(-\sum_{i=1}^{n} f_{i}(x_{i})  -\sum_{i=1}^{n} (\beta^{\ast}+q_{i}^{\ast})e_{i} \Big),\\
			\mathbf{y}^{\star} & \in \argmin_{\mathbf{y} \in (\mathbb{R}^{\geq 0})^{m}} \sum_{k=1}^{m}(\xi_{k}^{\ast} - \sum_{i=1}^{n}q_{i}^{\ast}\mathbf{A}_{ik})\mathbf{y}_{k}. \label{eq:y_min}
		\end{align}
	\end{subequations}
	Since~\eqref{eq:strong_duality_xe} is separable over $i \in \mathcal{N}$, an equivalent formulation is 
	$
	(x_{i}^{\star},e_{i}^{\star}) \in \argmin_{(x_{i}^{\star},e_{i}^{\star}) \in \mathbb{X}_{i}}\Big(  - f_{i}(x_{i})  - (\beta^{\ast}+q_{i}^{\ast})e_{i} \Big), \quad i \in \mathcal{N}.
	$
	From~\eqref{eq:y_min}, the stationarity condition for $\mathbf{y}_{k}, k = 1,\dots,m$ is that $\xi_{k}^{\ast} - \sum_{i=1}^{n}q_{i}^{\ast}\mathbf{A}_{ik}=0$. Therefore, $(x_{i}^{\star},e_{i}^{\star})$ solves the optimization problem \eqref{eq:ce_max} and it optimal duals satisfy Eqs.~\eqref{eq:ce_price}-\eqref{eq:ce_y_stationary}.
	
	(ii) The proof of this part can be obtained by reversing the proof of part (i). $\hfill \square$ 
	
	\begin{proposition}
		Consider a TMAS-FN. Suppose each $f_{i}(\cdot)$ is concave over the domain $\mathbb{R}^{\geq 0}$. Then the optimal trading price $\lambda_{i}^{\ast}$ at each agent $i \in \mathcal{N}$ under the competitive equilibrium is nonnegative, i.e., $\lambda_{i}^{\ast} \geq 0.$
	\end{proposition}
	{\it \noindent Proof.}  Assume $\lambda_{i}^\ast<0$. Then $f_i(x_i) + \lambda_{i}^\ast e_i$ is a strictly decreasing function with respect to $e_i$. Since $e_i$ is unbounded below and upper bounded by $e_i \leq a_i-x_i$ in~\eqref{eq:ce_max}, there can not be a finite ${e}^\ast_i$ such that  ${e}^\ast_i\in \arg \max_{e_i\in \mathbb{R}}\big( f_i(x_i) + \lambda_{i}^\ast e_i\big)$, contradicting the definition of the competitive equilibrium. This completes the proof.  $\hfill \square$

	
	\begin{theorem}\label{thm:unique_e}
		Consider a TMAS-FN with infinite arc capacity $\mathbf{u}_{k} = \infty, k = 1,\dots,m$. Suppose each $f_{i}(\cdot)$ is strictly concave over the domain $\mathbb{R}^{\geq 0}$.  Suppose there exists a competitive equilibrium with positive trading prices $\lambda_{1}^{\ast}, \lambda_{2}^{\ast}, \dots, \lambda_{n}^{\ast} > 0$. Then the trading decisions $e_{i}^{\ast}, i \in \mathcal{N}$ are unique. 
	\end{theorem}
	{\it Proof.} Let $(\beta^{\star}, \bm{\lambda}^{\star}, \mathbf{q}^{\star},\bm{\xi}^{\star})$ be optimal duals of~\eqref{eq:swe}. For each $i \in \mathcal{N}$, it can be derived that $\lambda^{\star} = - ( \beta^{\star} + q_{i}^{\star}) > 0$, which is equivalent to positive $\lambda_{i}^{\ast} >0$ in~\eqref{eq:ce_price} due to the equivalence of social welfare equilibrium and competitive equilibrium. Inequality constraint~\eqref{eq:swe_e_x_a} becomes the equality constraint $e_{i} = a_{i} -x_{i}, i \in \mathcal{N}$ due to complementary slackness conditions. Combining~\eqref{eq:swe_balance}, these two  equality constraints becomes one equality $\sum_{i  \in \mathcal{N}}x_{i} = \sum_{i  \in \mathcal{N}}a_{i}$. Furthermore, infinite arc capacity means inactive capacity constraint~\eqref{eq:swe_capacity_constraints}. Therefore, the optimization problem~\eqref{eq:swe} reduces to
	\begin{subequations}\label{eq:swe_degenerated}
		\begin{align}
			\max_{\mathbf{x}} \quad &\sum_{i  \in \mathcal{N}}f_{i}(x_{i}) \\
			{\text{\rm s.t.}} \quad &\sum_{i  \in \mathcal{N}}x_{i} = \sum_{i  \in \mathcal{N}}a_{i},\\
			\quad & x_{i} \geq 0, i \in \mathcal{N}.
		\end{align}
	\end{subequations}
	Since $f_{i}(x_{i})$ is strictly concave in $x_{i}$, the optimal solution  $\mathbf{x}^{\star}$ for~\eqref{eq:swe_degenerated} is unique. Therefore, according to $e_{i} = a_{i} -x_{i}, i \in \mathcal{N}$, each optimal $e_{i}^{\star}$ can be uniquely computed. The proof is now completed. However, note that the flow balance equality constraints in~\eqref{eq:swe_flow} contain $n$ linear equations regarding $m$ variables of $\mathbf{y}_{k}^{\star}, k = 1, \dots, m$. Even by combining inequality constraints~\eqref{eq:swe_positive_y}, the optimal flow $\mathbf{y}^{\star}$ might not be unique. $\hfill \square$
	
	\section{Social Acceptance: Homogeneous Pricing }\label{sec:homogeneous_pricing}
	In this section, we investigate the fair case in which all agents have the same trading prices.
	\subsection{Equal Prices Condition}
	We first present the notion of the standard social welfare equilibrium and the standard competitive equilibrium for standard MAS in the absence of network flow constraints. In other words, agents can trade with each other through the underlying network freely.
	\begin{definition}[Standard Social Welfare Equilibrium]\label{def:standard_swe}
		A standard social welfare equilibrium $(\mathbf{x}^{\star},\mathbf{e}^{\star})$ is achieved for a standard MAS if $(\mathbf{x}^{\star},\mathbf{e}^{\star})$  solves the following problem:
		\begin{subequations}\label{eq:standard_swe}
			\begin{align}
				\max_{\mathbf{x},\mathbf{e}} \quad & \sum_{i=0}^n f_i(x_i)\label{eq11} \\
				{\rm s.t.} 	\quad  & \sum_{i=0}^n e_i = 0,\label{eq12}\\
				\quad & e_i\leq a_i - x_i, \  i =1,\dots,n,\label{eq13} \\
				\quad & x_i\in \mathbb{R}^{\geq0}, \ i=1,\dots,n.\label{eq14}
			\end{align}
		\end{subequations}
	\end{definition}
	\begin{definition}[Standard Competitive Equilibrium]
		A standard competitive equilibrium ($\mathbf{x}^\ast$, $\mathbf{e}^\ast$, $\lambda_{0}^\ast$) is achieved for a standard MAS if the following conditions hold.
		
		(i) For each agent $i \in \mathcal{N}$, the pair $(\mathbf{x}^\ast$, $\mathbf{e}^\ast)$ is an optimizer to the following maximization problem:
		\begin{equation}
			\begin{aligned}
				\max_{{x}_i, e_i} \quad &  f_i(x_i)+\lambda_{0}^\ast e_i \\
				{\rm s.t.} \quad & x_i+e_i\leq a_i, \\
				\quad & x_i\in \mathbb{R}^{\geq0}, e_i\in \mathbb{R}.
			\end{aligned}\label{trading_ceq}
		\end{equation}
		
		(ii) The total demand and supply are balanced across the network; that is,
		\begin{equation}\label{trading_demand_supply_constraints}
			\sum_{i=0}^n e_i^\ast =0.
		\end{equation} 
	\end{definition}
	The equivalence of the standard social welfare equilibrium and the standard competitive equilibrium is proved \cite{chen2022competitive}.
	\medskip
	
	We next present a special case where the optimal trading prices at all agents are equal to each other.
	\begin{definition}[Interior Implementation]
		For a transactive multi-agent system, the optimal trading decision $(e_{1}^{\star},\dots,e_{n}^{\star})$ under the standard social welfare equilibrium can be interiorly implemented over a flow network if there exist flows $\mathbf{y}_{k}, k= 1,2,\dots,m$ such that
		\begin{subequations}\label{eq:interior_condition}
			\begin{align}
				\quad &e_{i}^{\star} = \sum_{k = 1}^{m} \mathbf{A}_{ik}\mathbf{y}_{k}^{\star},  \  i = 1,\dots,n, \label{eq:interior_e_y}\\
				\quad & 0<\mathbf{y}_{k}^{\star} < \mathbf{u}_{k}, \ k = 1,\dots,m. \label{eq:interior_y_range}
			\end{align}
		\end{subequations}
	\end{definition}
	
	
	\begin{theorem}\label{thm:interior_imple}
		Suppose that trading decisions $(e_{1}^{\star},\dots,e_{n}^{\star})$ in a standard social welfare equilibrium can be interiorly realized over an FN. Then $(e_{1}^{\star},\dots,e_{n}^{\star})$ is a trading decision of a social welfare equilibrium for this TMAS-FN with equal trading prices at all agents, i.e., $\lambda_{1}^{\ast} = \lambda_{2}^{\ast}=\cdots=\lambda_{n}^{\ast}$.
	\end{theorem}
	{\it Proof.} Since $\mathbf{y}_{k}^{\ast} < \mathbf{u}_{k}$ for each $ k = 1,\dots,m$, Eq.~\eqref{eq:ce_capacity_dual} leads to $\xi_{k}^{\ast} = 0$ under the competitive equilibrium. Thus, Eq.~\eqref{eq:ce_y_stationary} becomes $\sum_{i=1}^{n}q_{i}^{\ast}\mathbf{A}_{ik}=0$, which is a system of $m$ linear equations with $n$ variables of $q_{i}, i \in \mathcal{N}$. For each  column $k$ of $\mathbf{A}$, corresponding to arc $(s,t)_{s,t \in \mathcal{N}} \in \mathcal{A}$, we have two nonzero entries $\mathbf{A}_{sk} = 1$ and $\mathbf{A}_{tk} = -1$, leading to $q_{s}^{\ast} = q_{t}^{\ast}.$ Since the underlying flow network is a connected graph, it follows that $q_{s}^{\ast} = q_{t}^{\ast}$ for all $s,t \in \mathcal{N}$. According to to~\eqref{eq:ce_price}, the trading prices of all agents are the same. $\hfill \square$
	
	\subsection{Social Acceptance Sharing}
	The optimal trading prices of agents, as optimal duals corresponding to~\eqref{eq:swe_e_x_a}, may be different for each agent. This is because the optimal trading price at agent $i \in \mathcal{N}$ is affected by the capacity of arcs corresponding to agent $i$. Given the local resources $\mathbf{a}$, arc capacity $\mathbf{u}$, and node-arc incidence matrix $\mathbf{A}$, the optimal trading prices rely on the utility functions of agents. Without restrictions on the choice of utility functions, the optimal trading prices may be different, which creates unfairness for resource allocation in the TMAS-FN and disadvantages certain agents. To improve the fairness of trading prices, we need an approach that leads to equal trading prices across the network. In this section, we define a social acceptance sharing problem in the TMAS-FN, and present a symbolic algorithm regarding how the social acceptance sharing problem can be solved conceptually.
	\medskip
	
	{\noindent \bf Social Acceptance Sharing Problem for TMAS-FN:} Conisder a TMAS-FN whose agents $i \in \mathcal{N}$ have parameterized concave utility functions $f(\cdot;\bm{\theta}_{i})$ with $\bm{\theta}_{i} \in \mathbb{R}^{p}$. A utility function is said to be socially admissible if for agent $i \in \mathcal{N}$, the $j$th parameter of $\bm{\theta}_{i}$ satisfies  $\theta_{i}^{[j]} \in [\theta_{\min}^{[j]},\theta_{\max}^{[j]}], j \in \mathcal{P} := \{1,\dots,p\}$.   Find a range $\Theta $ for $(\theta_{\min}^{[j]},\theta_{\max}^{[j]})_{j \in \mathcal{P}}$ such that if $\theta_{i}^{[j]} \in [\theta_{\min}^{[j]},\theta_{\max}^{[j]}], j \in\mathcal{P}$, $i \in \mathcal{N}$, then it yields equal prices at all agents, i.e., $\lambda_{1}^{\ast} = \lambda_{2}^{\ast} = \dots = \lambda_{n}^{\ast} >0$ under the competitive equilibrium.
	
	We now present Algorithm~\ref{alg:sym_Theta} for conceptual computation of $\Theta$.
	\begin{algorithm}
		\caption{Conceptual Computation for $\Theta$}
		\label{alg:sym_Theta}
		
		\begin{algorithmic}[1]
			\State Compute $\mathscr{E}(\bm{\theta}) = \{\bf{e}^{\star}|(\bf{x}^{\star},\bf{e}^{\star}) \text{ solves \eqref{eq:standard_swe}}\}$.
			
			\State Define the set $\mathscr{K}(\theta_{\min}^{[1]},\theta_{\max}^{[1]},\dots,\theta_{\min}^{[p]},\theta_{\max}^{[p]}):=\cup_{\theta_{i}^{[j]} \in [\theta_{\min}^{[j]},\theta_{\max}^{[j]}], i \in \mathcal{N}, j\in \mathcal{P}} \mathscr{E}(\bm{\theta})$. 
			
			\State Define the set $\mathscr{M}:=\{\mathbf{h} \in \mathbb{R}^{n}|h = \mathbf{A}\mathbf{y}, 0\leq \mathbf{y}_{k} \leq \mathbf{u}_{k}, k = 1, \dots,m\}$.
			
			\State Compute $\Theta =\{(\theta_{\min}^{[1]},\theta_{\max}^{[1]},\dots,\theta_{\min}^{[p]},\theta_{\max}^{[p]})|\mathscr{K} \subseteq \mathscr{M}\}$.
		\end{algorithmic}
	\end{algorithm}
	
	Step 1 involves solving the optimization problem \eqref{eq:standard_swe} for given individual parameters $\bm{\theta}$, and recording the resulting set of optimal solutions  $\bf{e}^{\star}$. The notation $\mathscr{E}(\bm{\theta})$ refers to the set of all such optimal solutions $\bf{e}^{\star}$.
	Since each parameter $\theta^{[j]}_{i}$ for agent $i$ takes value from $[\theta_{\min}^{[j]},\theta_{\max}^{[j]}]$, Step 2 defines all optimal solutions of $\mathbf{e}^{\star}$ under all possible parameter configurations $\theta^{[j]}_{i} \in [\theta_{\min}^{[j]},\theta_{\max}^{[j]}], \forall j \in \mathcal{P}$ by set $\mathscr{K}$ depending on $(\theta_{\min}^{[1]},\theta_{\max}^{[1]},\dots,\theta_{\min}^{[p]},\theta_{\max}^{[p]})$. Step 3 defines set $\mathscr{K}$ as the image space of arc-node incidence matrix $\mathbf{A}$ subject to the constraints $0\leq \mathbf{y}_{k} \leq \mathbf{u}_{k}, k = 1, \dots,m$. Step 4 computes the range $\Theta$ for $(\theta_{\min}^{[1]},\theta_{\max}^{[1]},\dots,\theta_{\min}^{[p]},\theta_{\max}^{[p]})$ such that $\mathscr{K}$ is a subset of $\mathscr{M}$.
	
	\begin{remark}
		It should be noted that while this algorithm outlines how to compute the parameter range $\Theta$, it is not necessarily numerically implementable. However, it highlights the need for a conceptual framework to tackle the social acceptance sharing problem for TMAS-FN.
	\end{remark}
	\subsection{Linear-quadratic TMAS-FN over Undirected Star Graphs}
	In what follows, we consider linear-quadratic TMAS-FN over undirected star graphs where utility functions are in the linear-quadratic form. We present an approach to achieve social acceptance of sharing (equal optimal trading prices) under the competitive equilibrium by synthesizing a class of utility functions from which agents can select. We make the following assumption.
	\begin{assumption}\label{apt:f_lq}
		For $i \in \mathcal{N}$, let $f_{i}(x_{i}) = -\frac{1}{2}\theta^{[1]}_{i}x_{i}^{2} + \theta^{[2]}_{i}x_{i} $, where $\theta^{[1]}_{i}\in\mathbb{R}^{> 0}$ and $\theta^{[2]}_{i}\in\mathbb{R}^{\geq 0}$. A utility function $f_i$ is socially admissible  if   there hold $\theta^{[1]}_{i}\in[\theta^{[1]}_{\rm min},\theta^{[1]}_{\rm max}]$ and $\theta^{[2]}_i\in[\theta^{[2]}_{\rm min},\theta^{[2]}_{\rm max}]$.
	\end{assumption}
	
	\medskip
	
	
	\medskip
	
	{\noindent \bf Social Shaping Problem:} Consider a linear-quadratic TMAS-FN over an undirected star graph. Find the range for $\theta^{[1]}_{\rm min},\theta^{[1]}_{\rm max},\theta^{[2]}_{\rm min},\theta^{[2]}_{\rm max}$ under which  there always exists a competitive equilibrium that leads to equal positive prices $\lambda_{i}^{\ast} = \lambda_{j}^{\ast} > 0, \forall i,j \in \mathcal{N}$ for all socially admissible  utility functions. 
	\medskip
	
	We next present the following lemma and theorem.
	\begin{lemma}\label{eq:lemma_e}
		Consider a linear-quadratic MAS over an undirected star graph. If the optimal trading decision $\mathbf{e}^{\star}$ under the standard social welfare equilibrium is in
		\begin{align} \label{eq:e_flowbound_interior}
			\mathbb{E} = \{\mathbf{e}^{\star}|\mathbf{A}^{-}_{i}\mathbf{u}< e_{i}^{\star} <  \mathbf{A}^{+}_{i}\mathbf{u}, \forall i \in \mathcal{N}\},
		\end{align}
		then there exists flow vector $\mathbf{y}$ satisfying~\eqref{eq:interior_condition}.
	\end{lemma}
	{\it \noindent Proof.} Consider any $\mathbf{e}^{\star} \in \mathbb{E}$ under the standard social equilibrium.  We know from Eq.~\eqref{eq:e_flowbound_interior} that $- (\mathbf{u}_{1} + \dots + \mathbf{u}_{n-1}) < e_{1}^{\star} < (\mathbf{u}_{1} + \dots + \mathbf{u}_{n-1})$ and $-\mathbf{u}_{i-1} < e_{i}^{\star}<\mathbf{u}_{i-1}, i = 2, \dots, n$. Now, we try to find one optimal solution for $\mathbf{y}_{k}^{\rm op},k = 1, \dots, 2(n-1)$ which satisfies~\eqref{eq:interior_condition}. 
	
	Expanding~\eqref{eq:interior_e_y}, we obtain
	\begin{align*}
		e_{1}^{\star} &= \sum_{k=1}^{n-1}(\mathbf{y}_{k}^{\star} - \mathbf{y}_{n-1+k}^{\star}),\\
		e_{i+1}^{\star} & = -(\mathbf{y}_{i}^{\star} - \mathbf{y}_{n-1+i}^{\star}), \ i = 1,\dots,n-1.
	\end{align*}
	Since $\mathbf{y}_{k}^{\star}, k=1,\dots,m$ are independent and satisfy~\eqref{eq:interior_condition}, we can regard $\mathbf{y}_{k}^{\star} - \mathbf{y}_{n-1+k}^{\star}, k =1, \dots, n-1$ as new variables, whose range are $[-\mathbf{u}_{k}, \mathbf{u}_{k}], k = 1, \dots, n-1.$ We can always find $\mathbf{y}_{k}^{\rm op}$ and $\mathbf{y}_{n-1+k}^{\rm op}$ satisfying $\mathbf{y}_{k}^{\rm op} - \mathbf{y}_{n-1+k}^{\rm op} \in [-\mathbf{u}_{k}, \mathbf{u}_{k}]$. We now let $\mathbf{y}_{k}^{\rm op} - \mathbf{y}_{n-1+k}^{\rm op}, k =1,\dots,n-1$ take values of $-e^{\star}_{k+1}, k =1,\dots,n-1$. Then, it is true that $	e_{1}^{\star} = -\sum_{k = 1}^{n-1} e^{\star}_{k+1} = \sum_{k = 1}^{n-1}(\mathbf{y}_{k}^{\rm op} - \mathbf{y}_{n-1+k}^{\rm op})$. Therefore, we have found one optimal solution for $\mathbf{y}_{k}^{\rm op},k = 1, \dots, 2(n-1)$ satisfying~\eqref{eq:interior_condition}. $\hfill \square$
	
	\begin{theorem}\label{thm:equal_price}
		Consider a linear-quadratic MAS over an undirected star graph whose agents have limited local resources $a_{i} < \mathbf{A}^{+}_{i}\mathbf{u}, i \in \mathcal{N}$. Then the optimal trading prices of all agents   under the competitive equilibrium are positive and equal for all socially admissible utility functions as long as $\big(\theta^{[1]}_{\rm min}, \theta^{[1]}_{\rm max}, \theta^{[2]}_{\rm min}, \theta^{[2]}_{\rm max}\big)\in \mathscr{S}_\ast$ defined by 
		\begin{equation}\label{eq:setS}
			\begin{aligned}
				\mathscr{S}_\ast:= & \Big\{ \big(\theta^{[1]}_{\rm min}, \theta^{[1]}_{\rm max}, \theta^{[2]}_{\rm min}, \theta^{[2]}_{\rm max}\big)\in \mathbb{R}_{\geq 0}^4: \quad \\
				& \underbrace{\frac{\theta_{\rm min}^{[2]}}{\theta_{\rm max}^{[1]} } > \frac{C}{n}}_{\text{\rm condition 1}};
				\underbrace{\mathbf{A}_{i}^{-}\mathbf{u} \leq a_{i} - \frac{\theta_{\max}^{[2]}}{\theta_{\min}^{[1]}}, \forall i \in \mathcal{N}}_{\text{\rm condition 2}} \Big\}
			\end{aligned}
		\end{equation}
	\end{theorem}
	
	{\noindent \it Proof.} Under the standard competitive equilibrium, each agent $i$'s unique optimal consumption decision $x_{i}$ and optimal trading decision $e_{i}$ can be written as $x_{i}^{\ast} = \max\{0,  \frac{\theta_{i}^{[2]}- \lambda_{0}^{\ast}}{\theta_{i}^{[1]} }\}$ and $e_{i}^{\ast} = \min \{a_{i} - \frac{\theta_{i}^{[2]}- \lambda_{0}^{\ast}}{\theta_{i}^{[1]} }, a_{i}\}$ with $\lambda_{0}^{\ast} = \frac{\sum_{i=1}^{n}\frac{\theta_{i}^{[2]}}{\theta_{i}^{[1]}} -C}{\sum_{i=1}^{n}\frac{1}{\theta_{i}^{[1]}}}.$ 
	From condition 1 in~\eqref{eq:setS}, the optimal price $\lambda_{0}^{\ast}$ is ensured to be positive because $ \sum_{i=1}^{n}\frac{\theta_{\min}^{[2]}}{\theta_{\max}^{[1]}} -C \geq n \frac{\theta_{\rm min}^{[2]}}{\theta_{\rm max}^{[1]} } -C > 0.$ From condition 2 in~\eqref{eq:setS}, the optimal trading decision $e_{i}^{\ast}$ is lower bounded such that $e_{i}^{\ast} > a_{i} - \frac{\theta_{i}^{[2]}}{\theta_{i}^{[1]} } \geq a_{i}- \frac{\theta_{\max}^{[2]}}{\theta_{\min}^{[1]} } \geq \mathbf{A}_{i}^{-}\mathbf{u}$. The $e_{i}^{\ast}$ is upper bounded such that $e_{i} \leq a_{i} \leq \mathbf{A}_{i}^{+}\mathbf{u}.$ Then, Lemma~\ref{eq:lemma_e} and Theorem~\ref{thm:interior_imple} hold. We finally obtain that $\lambda_{i} = \lambda_{j}, \forall i,j \in \mathcal{N}$ for all socially admissible utility functions with $\big(\theta^{[1]}_{\rm min}, \theta^{[1]}_{\rm max}, \theta^{[2]}_{\rm min}, \theta^{[2]}_{\rm max}\big)\in \mathscr{S}_\ast$. This completes our proof. $\hfill \square$
	\section{Numerical Examples}\label{sec:examples}
	In the following experiments, we consider transactive multi-agent systems over undirected flow networks with $20$ nodes and $30$ undirected edges ($60$ directed arcs). Although these flow networks are undirected, the underlying node-arc matrices still have $20$ rows and $60$ columns. These flow networks are generated based on Erd\H{o}s-R\'{e}nyi (ER) model. We ensure the connectivity of these flow networks by keeping generating a new ER network until it is connected.
	\medskip
	
	\begin{experiment}[Equivalence between CE and SWE]\label{exp1}
		Consider a transactive multi-agent system over a flow network with $20$ agents. There are $60$ directed arcs ($30$ undirected linkages) between agents. Each agent $i$
		is associated with a utility function $f_{i}$, which is represented by $f_{i}(x_{i}) = -\frac{1}{2}\theta^{[1]}_{i}x_{i}^{2} + \theta^{[2]}_{i}x_{i}$. 
		\medskip
		
		{\noindent \bf System Setting.} Each arc $k$'s capacity $\mathbf{u}_{k}, k=1,\dots,60$ is randomly generated from the interval $[0,2]$.  Each agent $i$'s local resource $a_{i}$  is a random number in the interval $[0,5]$. Each $\theta_{i}^{[1]}$ is randomly generated from the interval $[0.5,0.6]$, whereas each $\theta_{i}^{[2]}$ is randomly produced from the interval $[20]$.
		\medskip
		
		{\noindent \bf Two System-level Equilibria.} The social welfare equilibrium $(\mathbf{x}^{\star},\mathbf{e}^{\star}, \mathbf{y}^{\star})$ can be computed by numerically  solving the optimization problem \eqref{eq:swe}. The corresponding optimal  dual variables associated with the trading supply and demand balance constraint \eqref{eq:swe_balance}, $\beta^{\star}$, associated with the flow supply and demand balance constraints \eqref{eq:swe_flow}, $\mathbf{q}^{\star}$, and associated with~\eqref{eq:swe_e_x_a}, $\bm{\lambda}^{\star}$,  can also be obtained. Letting $\bm{\lambda}^{\ast} = \bm{\lambda}^{\star}$, we then compute a competitive equilibrium $(\mathbf{x}^{\ast},\mathbf{e}^{\ast}, \mathbf{y}^{\ast})$  that satisfies~\eqref{eq:ce_max}-\eqref{eq:ce_y_stationary}. We plot optimal trading decisions $\mathbf{e}^{\star}$ and $\mathbf{e}^{\ast}$ under the social welfare equilibrium and the competitive equilibrium in Fig.~\ref{fig:swe_ce_e}, whereas optimal consumption decisions $\mathbf{x}^{\star}$ and $\mathbf{x}^{\ast}$ under the social welfare equilibrium and the competitive equilibrium are plotted in Fig.~\ref{fig:swe_ce_x}. Furthermore, optimal duals of $\bm{\lambda}^{\star}$ and $\mathbf{-q}^{\star}$ under the social welfare equilibrium are showed in Fig.~\ref{fig:duals}.
		\begin{figure}[tb]
			\centering
			\includegraphics[width=0.38\textwidth]{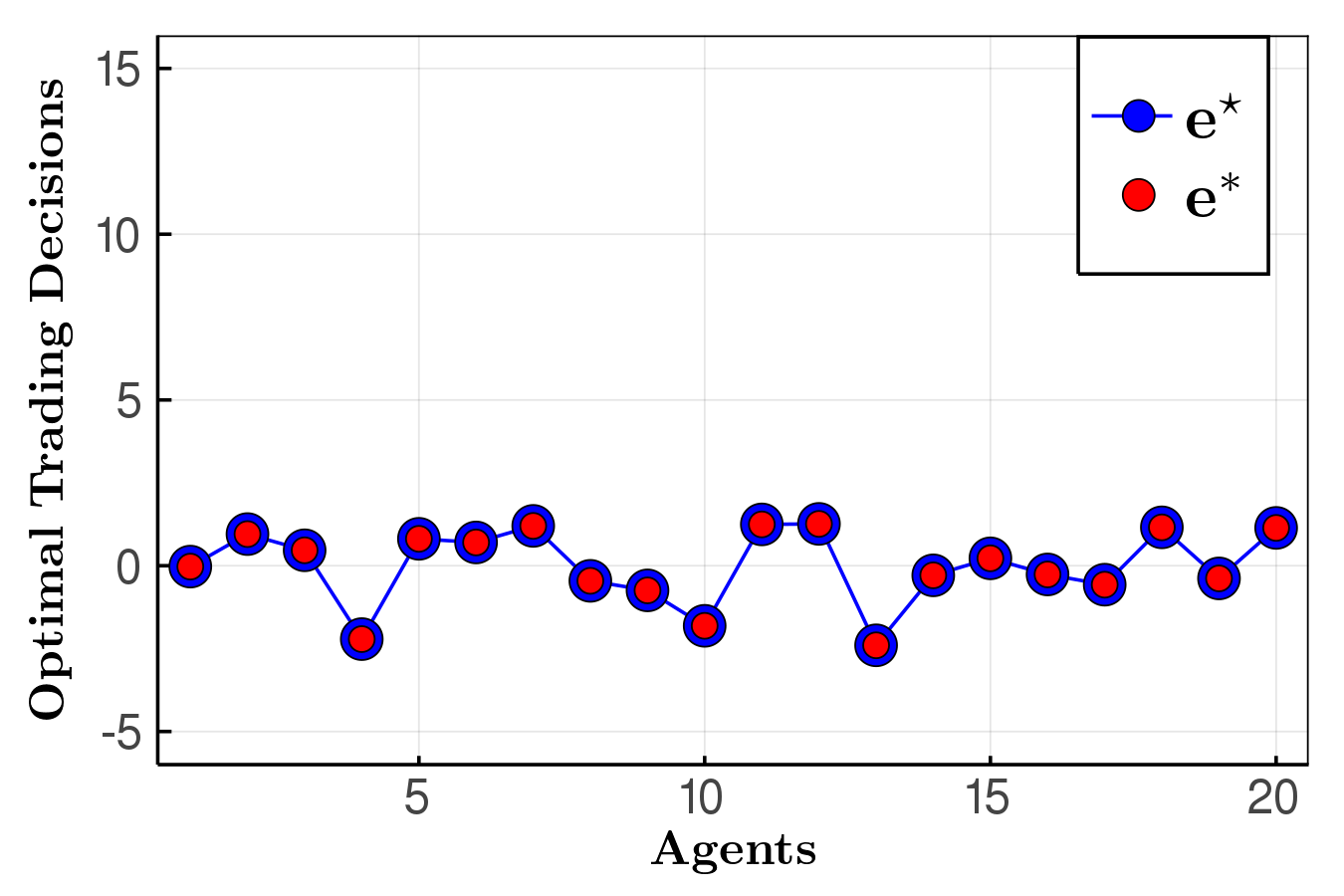}
			\caption{Optimal trading decisions $\mathbf{e}^{\star}$ and $\mathbf{e}^{\ast}$ under social welfare equilibrium and competitive equilibrium in Experiment~\ref{exp1}.} 
			\label{fig:swe_ce_e}
		\end{figure} 
		
		\begin{figure}[tb]
			\centering
			\includegraphics[width=0.38\textwidth]{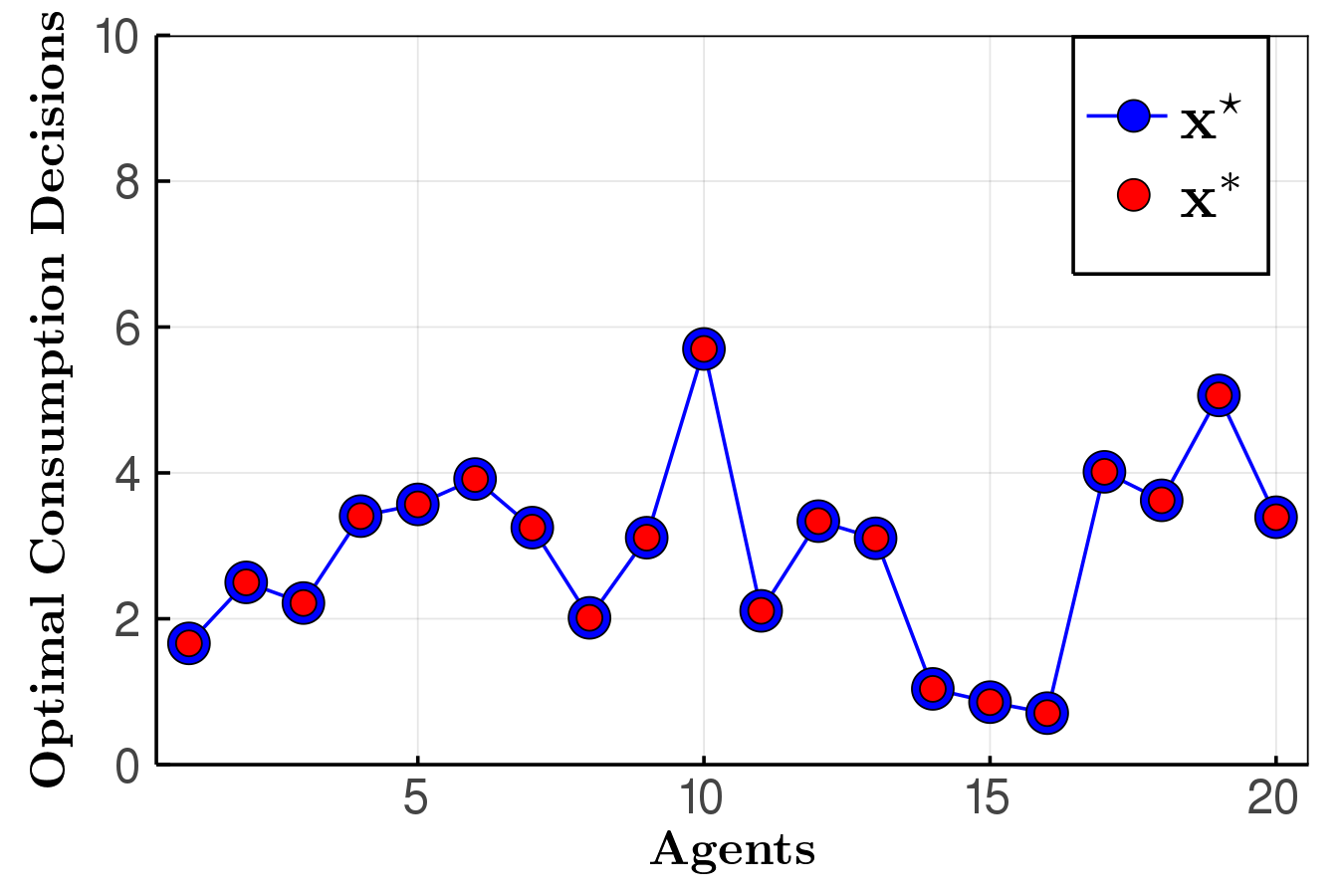}
			\caption{Optimal consumption decisions $\mathbf{x}^{\star}$ and $\mathbf{x}^{\ast}$ under social welfare equilibrium and competitive equilibrium in Experiment~\ref{exp1}.} 
			\label{fig:swe_ce_x}
		\end{figure} 
		
		\begin{figure}[tb]
			\centering
			\includegraphics[width=0.38\textwidth]{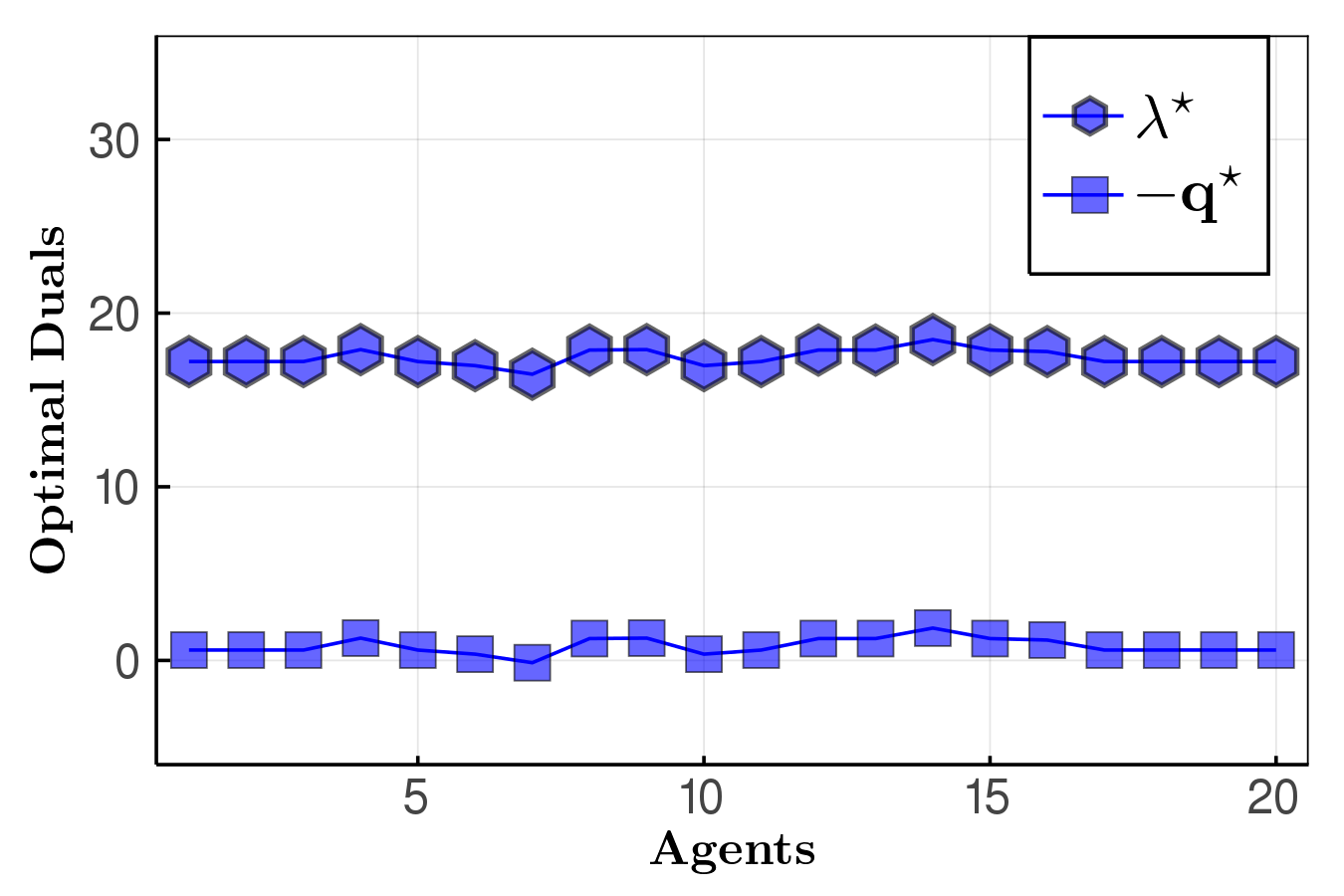}
			\caption{Optimal duals of $\bm{\lambda}^{\star}$ and $\mathbf{-q}^{\star}$ under social welfare equilibrium in Experiment~\ref{exp1}.} 
			\label{fig:duals}
		\end{figure} 
		
		From Figs.~\ref{fig:euqal_swe_ce_e} and~\ref{fig:euqal_swe_ce_x}, we observe that optimal trading decisions and optimal consumption decisions under the social welfare equilibrium and the dynamic competitive equilibrium agree. The optimal solutions for $\mathbf{y}^{\star}$ are not unique. However, since $\mathbf{e}^{\star}$ exists, the corresponding $\mathbf{y}^{\star}$ must obey ~\eqref{eq:swe_flow}. Consequently, we find  $\mathbf{y}^{\ast} = \mathbf{y}^{\star}$ as one optimal flow solution under the competitive equilibrium. The above analysis verifies Theorem~\ref{thm:equivalence}.
		We know that optimal duals of $\bm{\lambda}^{\star}$ and $\bm{q}^{\star}$ under the social welfare equilibrium are $\bm{\lambda}^{\ast}$ and $\bm{-q}^{\ast}$ under the competitive equilibrium. From Fig.~\ref{fig:duals}, it is noticeable that $\bm{\lambda}^{\star}$ is the translation of $\bm{-q}^{\star}$ along the vertical axis, which is consistent with the second condition~\eqref{eq:ce_price} in Definition~\ref{def:ce} of competitive equilibrium.$\hfill \square$
	\end{experiment}
	\medskip
	
	\begin{experiment}[Infinite Arc Capacity] \label{exp2}
		Consider the same transactive multi-agent system over a flow network in Experiment~\ref{exp1}. Use the same random seed in Experiment~\ref{exp1}'s system setting. We only modify values for $\mathbf{u}_{k}, k = 1,\dots,m$ to make them sufficiently large. 
		
		We compute the social welfare equilibrium $(\mathbf{x}^{\star},\mathbf{e}^{\star}, \mathbf{y}^{\star})$ for this TMAS-FN by  solving the optimization problem \eqref{eq:swe}. The corresponding optimal  dual variables associated  with~\eqref{eq:swe_e_x_a}, $\bm{\lambda}^{\star}$, and associated  with~\eqref{eq:swe_flow} are obtained. We compute the standard social welfare equilibrium $(\mathbf{x}^{\star}_{sd},\mathbf{e}^{\star}_{\rm sd})$ by  solving the optimization problem~\eqref{eq:standard_swe}. The corresponding optimal dual variables associated  with~\eqref{eq12}, $\bm{\lambda}^{\star}_{\rm sd}$, are obtained. We plot optimal trading decisions $\mathbf{e}^{\star}$ under the social welfare equilibrium and  $\mathbf{e}^{\star}_{\rm sd}$ under the standard social welfare equilibrium in Fig.~\ref{fig:euqal_swe_ce_e}. We plot optimal trading decisions $\mathbf{x}^{\star}$ under the social welfare equilibrium and  $\mathbf{x}^{\star}_{\rm sd}$ under the standard social welfare equilibrium in Fig.~\ref{fig:euqal_swe_ce_x}. We plot optimal duals of $\bm{\lambda}^{\star}$ under the social welfare equilibrium and $\bm{\lambda}^{\star}_{\rm sd}$ under the standard social welfare equilibrium in Fig.~\ref{fig:equal_duals}.
		\begin{figure}[tb]
			\centering
			\includegraphics[width=0.38\textwidth]{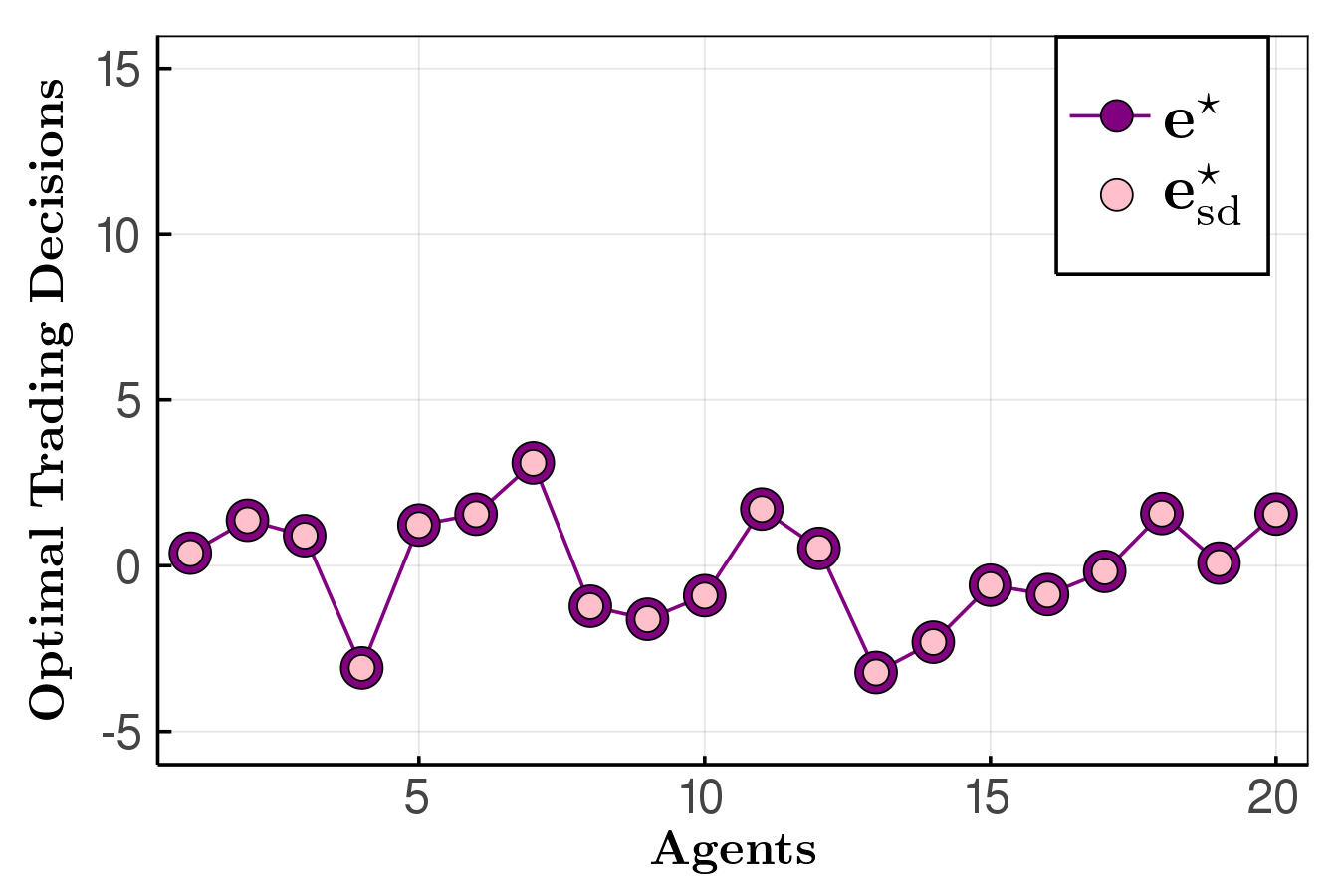}
			\caption{Optimal trading decisions $\mathbf{e}^{\star}$ under the social welfare equilibrium and  $\mathbf{e}^{\star}_{\rm sd}$ under  standard social welfare equilibrium in Experiment~\ref{exp2}.} 
			\label{fig:euqal_swe_ce_e}
		\end{figure} 
		
		\begin{figure}[tb]
			\centering
			\includegraphics[width=0.38\textwidth]{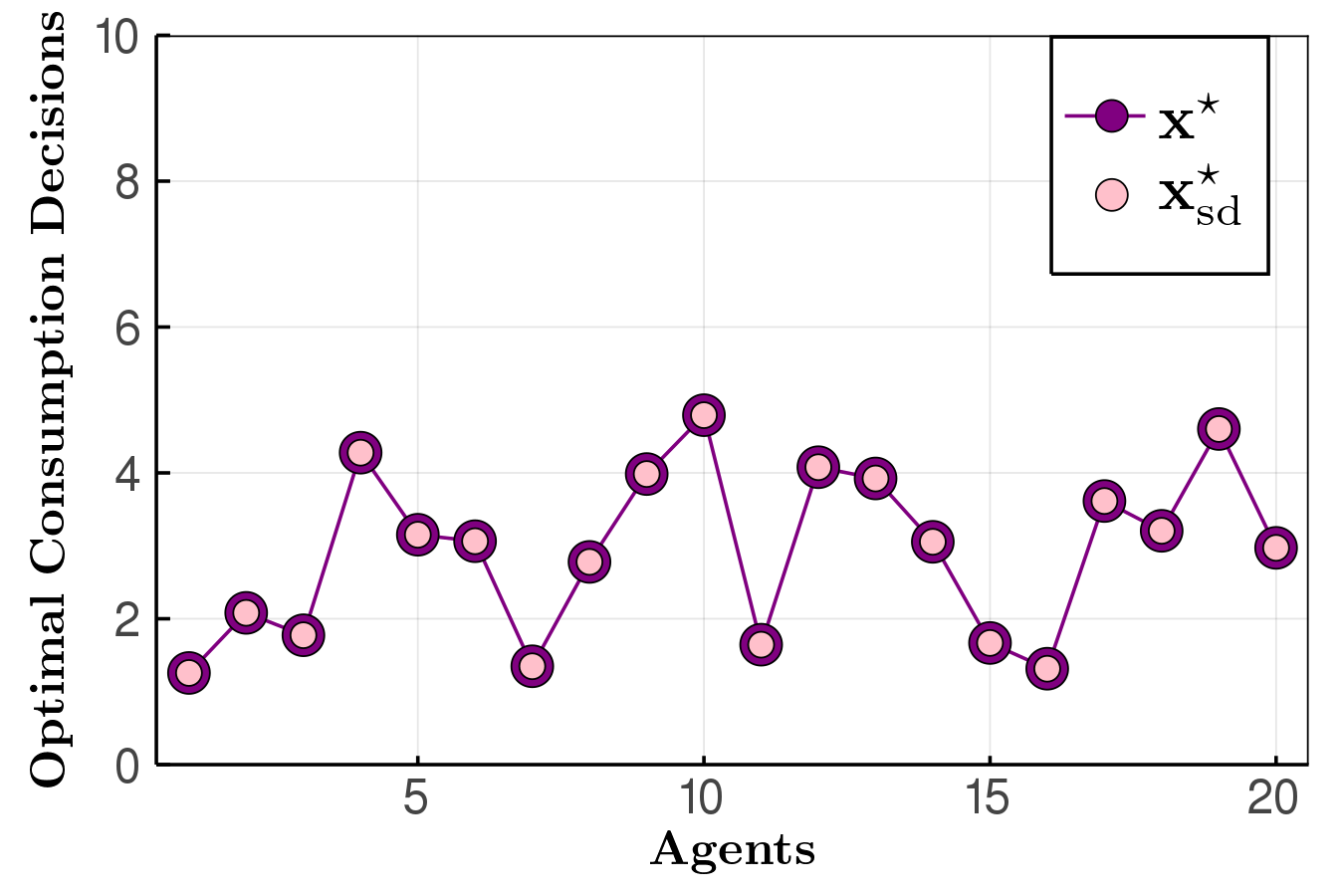}
			\caption{Optimal trading decisions $\mathbf{x}^{\star}$ under the social welfare equilibrium and  $\mathbf{x}^{\star}_{\rm sd}$ under  standard social welfare equilibrium in Experiment~\ref{exp2}.} 
			\label{fig:euqal_swe_ce_x}
		\end{figure} 
		
		\begin{figure}[tb]
			\centering
			\includegraphics[width=0.38\textwidth]{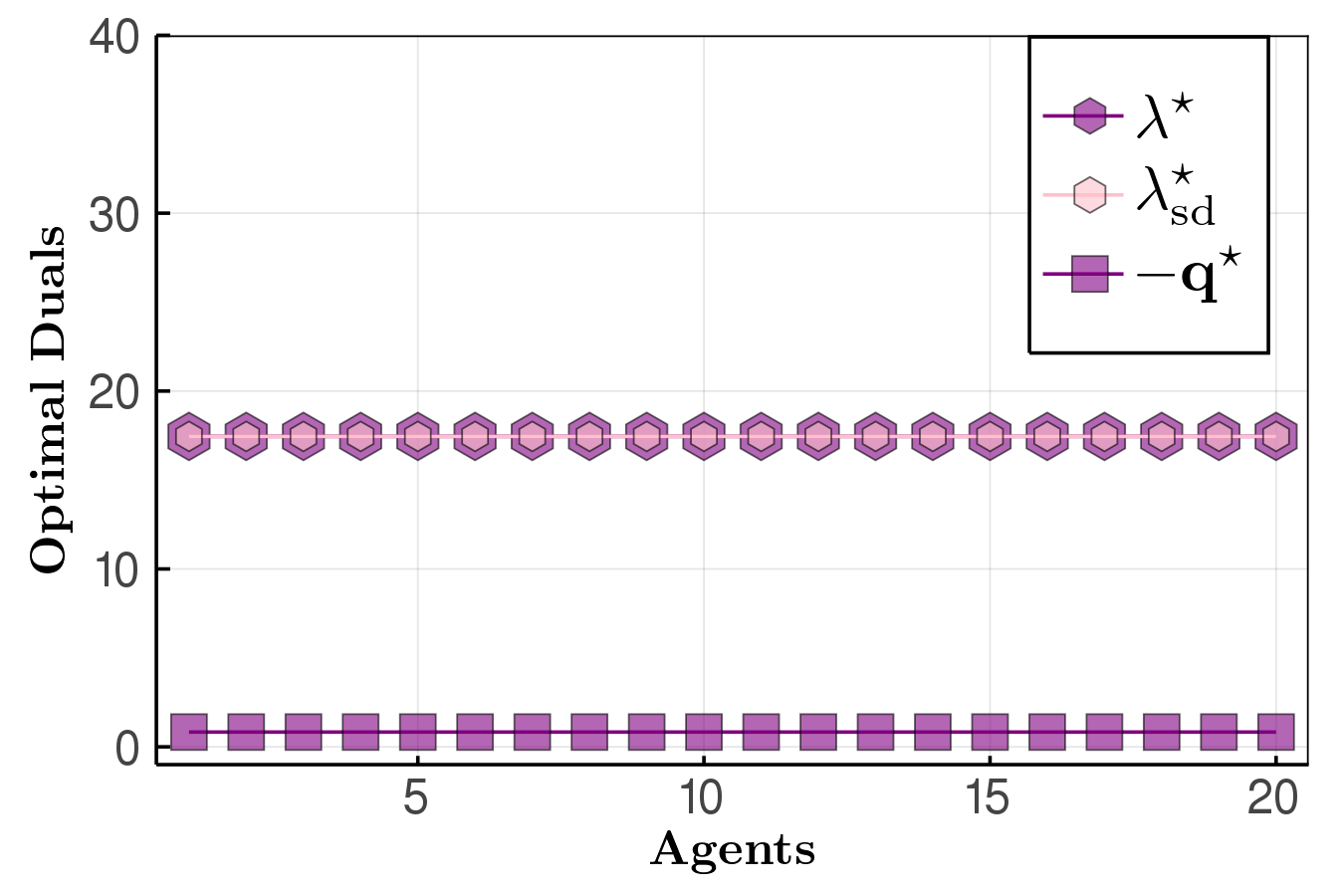}
			\caption{Optimal duals of $\bm{\lambda}^{\star}$ under social welfare equilibrium and $\bm{\lambda}^{\star}_{\rm sd}$ under standard social welfare equilibrium in Experiment~\ref{exp2}.} 
			\label{fig:equal_duals}
		\end{figure} 
		
		We can observe from Fig.~\ref{fig:equal_duals} that the  positive optimal trading prices $\bm{\lambda}^{\star}$ for this TMAS-FN with infinite arc capacity are equal to each other because the inequality capacity constraints~\eqref{eq:swe_capacity_constraints} are not activated, leading to all-zero $\mathbf{q}^{\star}$ (See~\eqref{eq:ce_y_stationary}). Furthermore, Figs.~\ref{fig:euqal_swe_ce_e} and \ref{fig:euqal_swe_ce_x} show the equivalence of ($\mathbf{x}^{\star},\mathbf{e}^{\star}$) and ($\mathbf{x}^{\star}_{\rm sd},\mathbf{e}^{\star}_{\rm sd}$), which is align with the proof of Theorem~\ref{thm:unique_e}. $\hfill \square$
	\end{experiment}
	\medskip

	\begin{experiment}[Trading Prices Vs. Arc Capacity]\label{exp3}
		Consider the same transactive multi-agent system over a flow network in Experiment~\ref{exp1}. Follow Experiment~\ref{exp1}'s system setting but with a different random seed. We only modify values for $\mathbf{u}_{k},k=1,\dots,m$, which are randomly generated from the interval $[0,\gamma]$. This parameter $\gamma$ is taken value from $\{0.01, 0.1, 0.5, 1, 10\}$. For each value of $\gamma$, we take a run to compute the social welfare equilibrium by numerically solving~\eqref{eq:swe}.
		
		We plot optimal duals of $\mathbf{q}^{\star}$ under the social welfare equilibrium versus parameter $\gamma$ in Fig.~\ref{fig:p_vs_u}. We plot optimal trading decisions $\mathbf{e}^{\star}$ under the social welfare equilibrium versus parameter $\gamma$ in Fig.~\ref{fig:e_vs_u}.  As indicated by the size of the circles, the smaller  the circles are, the smaller the value for $\gamma$ is. 
		\begin{figure}[tb]
			\centering
			\includegraphics[width=0.38\textwidth]{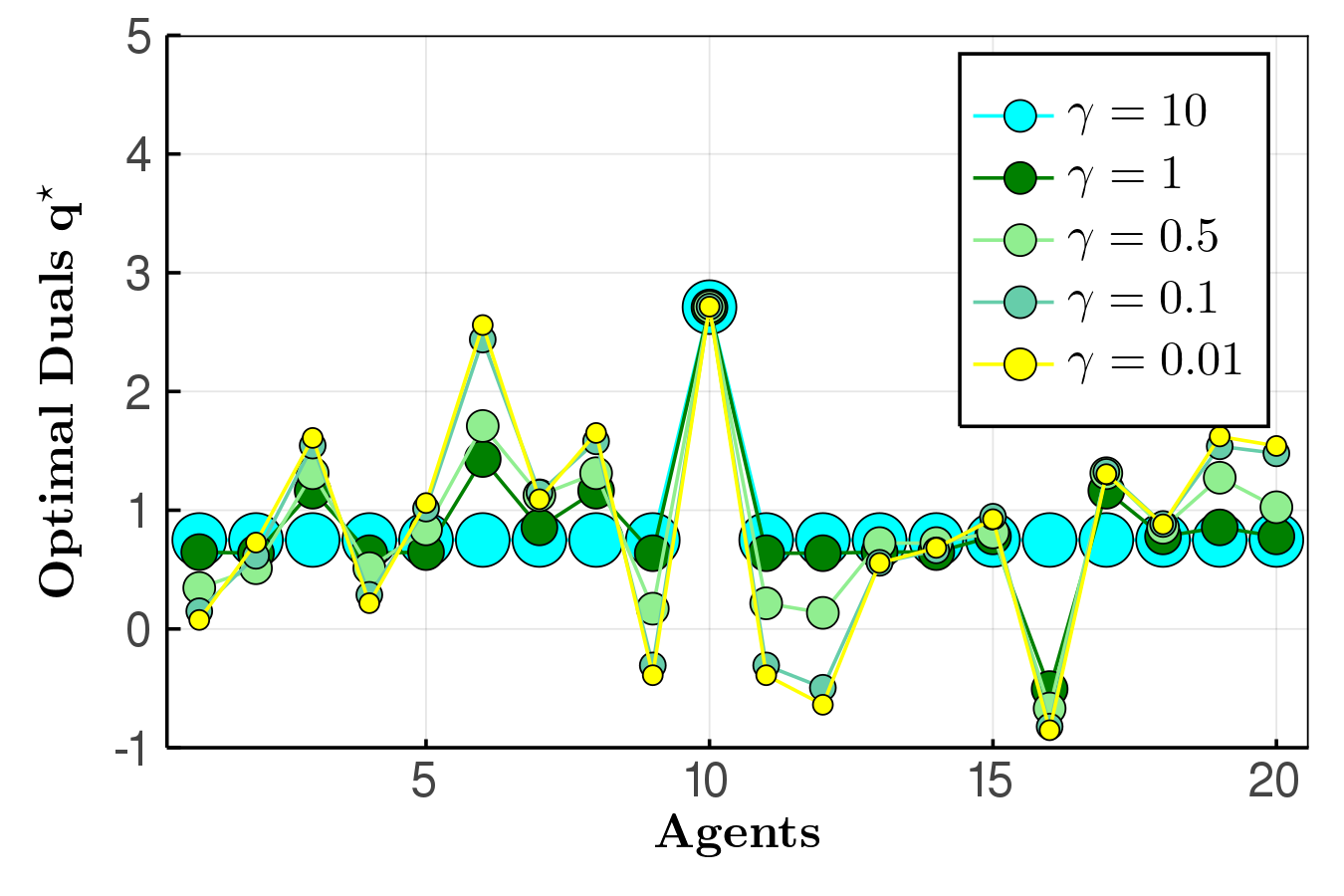}
			\caption{Optimal duals of $\mathbf{q}^{\star}$ under  social welfare equilibrium versus parameter $\gamma$ in Experiment~\ref{exp3}.} 
			\label{fig:p_vs_u}
			\vspace{-1em}
		\end{figure} 
		
		\begin{figure}[tb]
			\centering
			\includegraphics[width=0.38\textwidth]{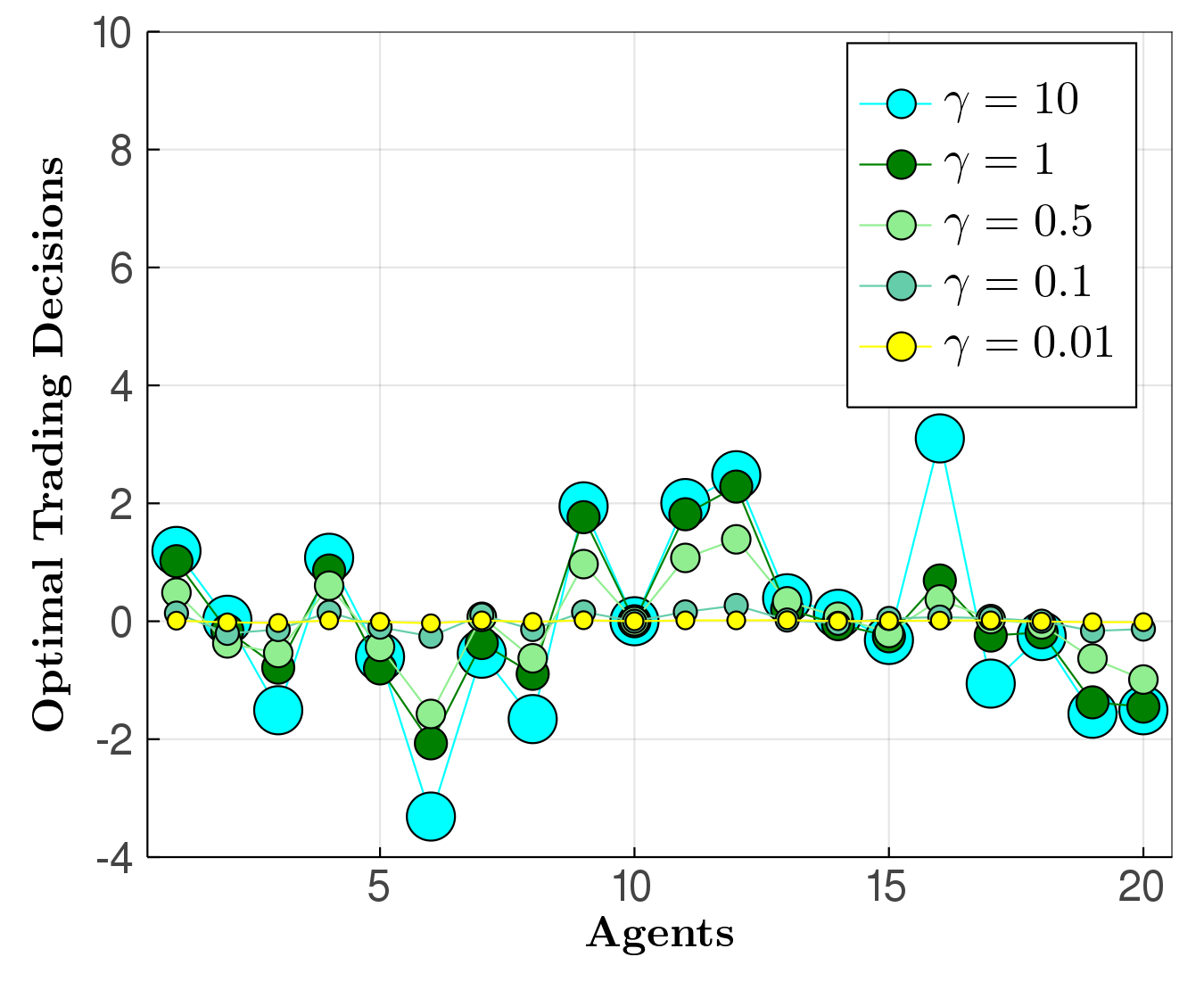}
			\caption{Optimal trading decisions under  social welfare equilibrium versus parameter $\gamma$ in Experiment~\ref{exp3}.} 
			\label{fig:e_vs_u}
			\vspace{-1em}
		\end{figure} 
		
		We can see from Fig.~\ref{fig:p_vs_u} that the optimal duals $q_{i}, i = 1,\dots,n$ become farther away from zero when $\gamma$ decreases. According to Eq.~\eqref{eq:ce_y_stationary}, when arc capacity becomes smaller, it is more likely for $\mathbf{y}^{\star}_{k}$ to reach the boundary $\mathbf{u}^{\star}_{k}$, inevitably leading to non-zero $\xi_{k}^{\star}$ as a punishment for preventing the breach of capacity constraints~\eqref{eq:swe_capacity_constraints}. As a consequence, more  $q_{i}$ will be driven to be non-zero to make Eq.~\eqref{eq:ce_capacity_dual} valid. It is shown in Fig.~\ref{fig:e_vs_u} that the optimal trading decisions $\mathbf{e}^{\star}$ become closer to zero when $\gamma$ decreases. This is mainly because the trading prices $\bm{\lambda}^{\star}$ become larger as a result of larger $\mathbf{q}$. Possibly, some agents gain more from consumption rather than trading. The trading activities over this TMAS-FN are thus discouraged to some extent. $\hfill \square$
	\end{experiment}
	\medskip
	
	\begin{experiment}\label{exp4}
		Consider a transactive multi-agent system over an undirected star graph with $5$ agents. There are $8$ directed arcs ($4$ undirected edges), whose capacity constraints are $\mathbf{u} = 15\cdot\mathbf{1}^{\top}_{8}$. Each agent has local resource $\mathbf{a} = 25\cdot\mathbf{1}^{\top}_{5}$. Each agent is associated with a linear-quadratic utility function as $f_{i}(x_{i}) = -\frac{1}{2}\theta^{[1]}_{i}x_{i}^{2} + \theta^{[2]}_{i}x_{i}$. Take $\theta_{\min}^{[1]} = 0.5$, $\theta_{\max}^{[1]} = 0.6$, $\theta_{\min}^{[2]} = 18$ and $\theta_{\max}^{[2]} = 20$. We can verify such a configuration of $\big(\theta^{[1]}_{\rm min}, \theta^{[1]}_{\rm max}, \theta^{[2]}_{\rm min}, \theta^{[2]}_{\rm max}\big)$ is a point in $\mathscr{S}_\ast$ defined in~\eqref{eq:setS}.
		
		We compute the social welfare equilibrium for this TMAS-FN  by  solving the optimization problem \eqref{eq:swe}. The corresponding optimal  dual variables associated  with~\eqref{eq:swe_e_x_a}, $\bm{\lambda}^{\star}$, are obtained as optimal trading prices, which are plotted in Fig.~\ref{fig:test_range}. We can observe that optimal trading prices at all agents are equal to each other. This provides validation for Theorem~\ref{thm:equal_price}. $\hfill \square$
		
		\begin{figure}[tb]
			\centering
			\includegraphics[width=0.38\textwidth]{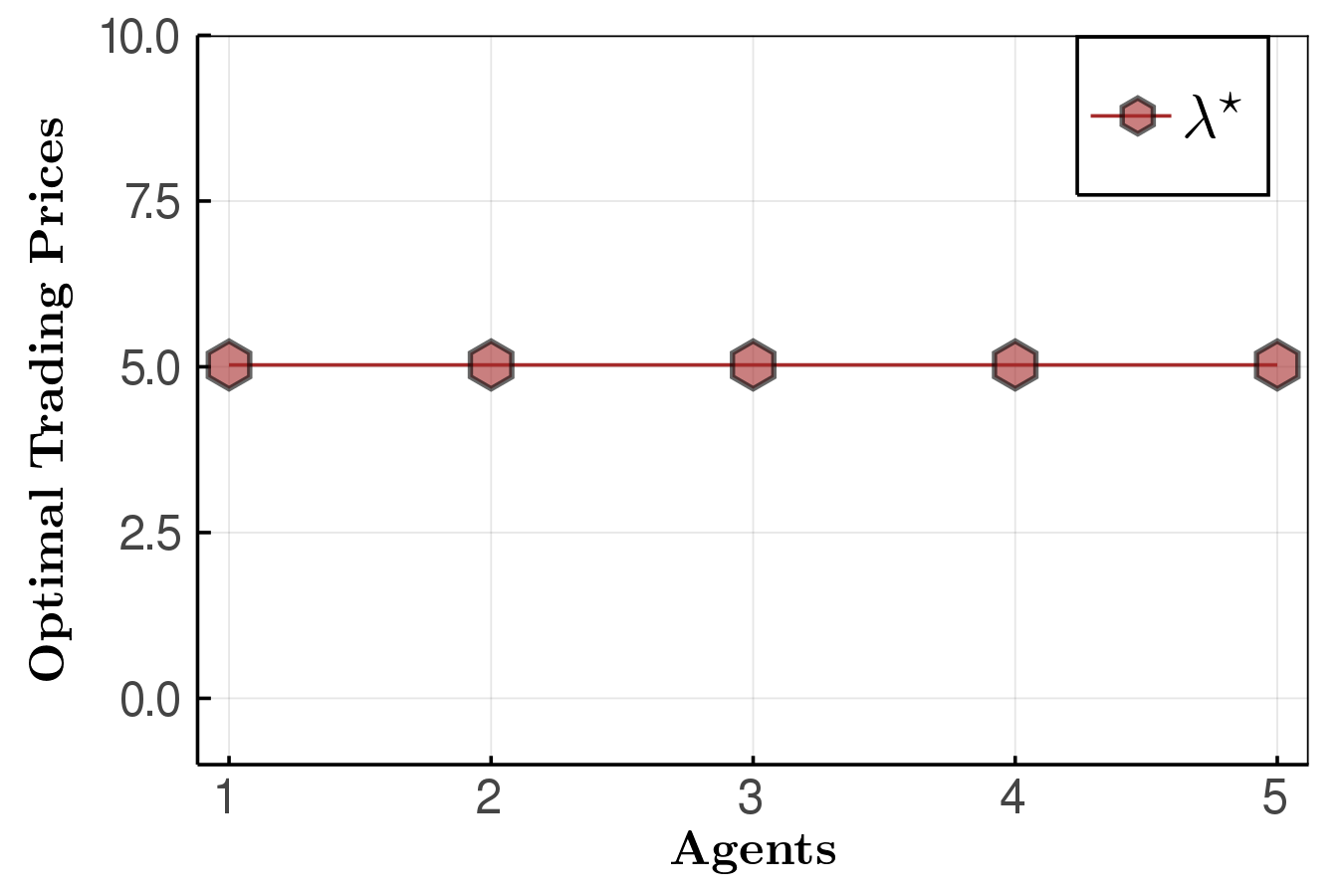}
			\caption{Equal optimal trading prices under  social welfare equilibrium when agents select their utility functions from a prescribed class of utility functions.} 
			\label{fig:test_range}
			\vspace{-1em}
		\end{figure} 
	\end{experiment}

	\section{Conclusion}\label{sec:conclusion}
	
	This paper presented insights into the implementation of transactive multi-agent systems over flow networks to facilitate the sharing of decentralized resources. We established a competitive market with a pricing mechanism that internalized sharing/flow capacity constraints. We demonstrated through duality theory that competitive equilibrium and social welfare equilibrium exist and coincide under convexity assumptions. We also defined a social acceptance sharing problem and proposed a conceptual computation method to solve the problem, with a special case of linear-quadratic MAS provided as an example. We validated the results through numerical extensive experiments. Future work to extend the work to the dynamic case is an interesting direction. Future work also involves explicitly constructing a range of socially admissible utility functions for generic flow networks to shape homogeneous pricing.
	
	\bibliographystyle{IEEEtran}
	\bibliography{reference}
\end{document}